\documentclass[conference]{IEEEtran}

\let\labelindent\relax
\usepackage[inline]{enumitem}  

\usepackage{booktabs}
\usepackage{multirow}                 
\usepackage{makecell}

\usepackage{graphicx}   
\usepackage{color}     
\usepackage[caption=false]{subfig}	 
\usepackage{pgfplots}	
\usetikzlibrary{patterns}        
\pgfplotsset{compat=1.14}

\usepackage{cite}               
\usepackage{amsthm}             
\usepackage{amsmath}
\usepackage{amssymb}
\usepackage{amsfonts}           

\usepackage{algorithm}      
\usepackage[noend]{algpseudocode}  

\newcommand{\system}{{\sc Templar}}

\newcommand{\eg}{e.g.}

\newcommand{\ie}{i.e.}

\newcommand{\etc}{etc.}

\newcommand{\naive}{na\"{i}ve}

\newcommand{\ignore}[1]{}

\newtheorem{example}{Example}

\newtheorem{definition}{Definition}

\algnewcommand{\IIf}[1]{\State\algorithmicif\ #1\ \algorithmicthen}
\algnewcommand{\EndIIf}{\unskip\ \algorithmicend\ \algorithmicif}

\begin{document}

\title{Bridging the Semantic Gap with SQL Query Logs in Natural Language Interfaces to Databases}

\author{\IEEEauthorblockN{Christopher Baik, H. V. Jagadish}
\IEEEauthorblockA{University of Michigan \\
Ann Arbor, MI, USA \\
Email: \{cjbaik, jag\}@umich.edu}
\and
\IEEEauthorblockN{Yunyao Li}
\IEEEauthorblockA{IBM Research - Almaden \\
San Jose, CA, USA \\
Email: yunyaoli@us.ibm.com}}

\maketitle

\begin{abstract}

A critical challenge in constructing a {\em natural language interface to database (NLIDB)} is bridging the semantic gap between a natural language query (NLQ) and the underlying data. Two specific ways this challenge exhibits itself is through {\em keyword mapping} and {\em join path inference}. {\em Keyword mapping} is the task of mapping individual keywords in the original NLQ to database elements (such as relations, attributes or values). It is challenging due to the ambiguity in mapping the user's mental model and diction to the schema definition and contents of the underlying database. {\em Join path inference} is the process of selecting the relations and join conditions in the \texttt{FROM} clause of the final SQL query, and is difficult because NLIDB users lack the knowledge of the database schema or SQL and therefore cannot explicitly specify the intermediate tables and joins needed to construct a final SQL query. In this paper, we propose leveraging information from the SQL query log of a database to enhance the performance of existing NLIDBs with respect to these challenges. We present a system \system\ that can be used to augment existing NLIDBs. Our extensive experimental evaluation demonstrates the effectiveness of our approach, leading up to 138\% improvement in top-1 accuracy in existing NLIDBs by leveraging SQL query log information.
\end{abstract}


\section{Introduction}


\begin{table*}[t]
  \centering
  \small
  \caption{State-of-the-art NLIDBs. Upper half are pipeline-based, lower half are end-to-end deep learning systems.}
  \begin{tabular}{lp{2.82cm}p{2.65cm}lp{2.8cm}l}
    \toprule
      \textbf{System} & \textbf{NLQ Pre-processing} & \textbf{Rel/Attr Mapping} & \textbf{Value Mapping} & \textbf{Join Path Inference} & \textbf{SQL Post-processing}\\
    \midrule
      Precise~\cite{popescu:precise} & \makecell[tl]{Tokenizer +\\Charniak~\cite{charniak2000maximum} parser} & WordNet~\cite{miller:wordnet} & Same as rel/attr & \makecell[tl]{Max-flow algorithm +\\User interaction} & N/A\\
      NaLIR~\cite{li:nalir} & Stanford Parser~\cite{demarneffe2006generating} & \makecell[tl]{WordNet~\cite{miller:wordnet} +\\User interaction} & Same as rel/attr & \makecell[tl]{Preset path weights +\\User interaction} & \makecell[tl]{Query tree heuristics +\\User interaction}\\
      SQLizer~\cite{yaghmazadeh:sqlizer} & Sempre~\cite{berant2013semantic} & word2vec~\cite{mikolov:word2vec} & Same as rel/attr & Hand-written repair rules & Hand-written repair rules\\
      ATHENA~\cite{saha2016athena} & Tokenizer & \makecell[tl]{Synonym lexicon +\\Pre-defined ontology} & \makecell[tl]{Index with\\semantic variants} & Pre-defined ontology & N/A\\
    \midrule
      Seq2SQL~\cite{zhong2017seq2sql} & \makecell[tl]{Tokenizer +\\Stanford CoreNLP~\cite{manning2014stanford}} & \makecell[tl]{GloVe~\cite{pennington2014glove} +\\character n-grams~\cite{hashimoto2016joint}} & Unsupported & N/A & N/A\\
      SQLNet~\cite{xu2017sqlnet} &  \makecell[tl]{Tokenizer +\\Stanford CoreNLP~\cite{manning2014stanford}} & GloVe~\cite{pennington2014glove} & Unsupported & N/A & N/A\\
      DBPal~\cite{utama2018end} & Replace literals with placeholders & Unspecified & word2vec~\cite{mikolov:word2vec} & \makecell[tl]{Select min-length\\ path} & \makecell[tl]{SQL syntax repair +\\Fill placeholders}\\
    \bottomrule
  \end{tabular}
  \vspace{-0.5cm}
  \label{tab:nlidb}
\end{table*}

From business intelligence to web design to answering everyday factual questions, today's technology-rich world is overwhelmingly dependent on database systems. End-users have various means to query these databases for information, whether through voice-based assistants such as Google Assistant\footnote{https://assistant.google.com/}, Apple Siri\footnote{https://www.apple.com/ios/siri}, or Amazon Echo\footnote{https://www.amazon.com/echo}; queries in a structured language such as SQL or LINQ; or keyword searches on search engines such as Google or Bing.

While relational databases have been around for decades, query languages to access such databases, such as SQL, are unlikely to ever become common knowledge for the average end-user. Accordingly, a major goal in this area is to provide a means for users to query databases using everyday language via a \textit{natural language interface to database (NLIDB)}.

The task of an NLIDB has been primarily modeled as the problem of translating a natural language query (NLQ) into a SQL query. State-of-the-art systems developed to solve this task take one of two architectural approaches:
\begin{enumerate*}[label=(\arabic*)]
\item the {\em pipeline approach} of converting an NLQ into intermediate representations then mapping these representations to SQL (e.g. ~\cite{popescu:precise,li:nalir,saha2016athena,yaghmazadeh:sqlizer}), and
\item the {\em deep learning approach} of using an end-to-end neural network to perform the translation (e.g. ~\cite{utama2018end,xu2017sqlnet,zhong2017seq2sql}) .
\end{enumerate*}

However, as pointed out by \cite{li2017natural}, one fundamental challenge in supporting NLIDBs is {\em bridging the semantic gap} between a NLQ and the underlying data. When translating NLQ to SQL, this challenge arises in two specific problems:
\begin{enumerate*}[label=(\arabic*)]
\item {\em keyword mapping} and
\item {\em join path inference}.
\end{enumerate*}
{\em Keyword mapping} is the task of mapping individual keywords in the original NLQ to database elements (such as relations, attributes or values). It is a challenging task because of the ambiguity in mapping the user's mental model and diction to the schema definition and contents of the database. {\em Join path inference} is the process of selecting the relations and join conditions in the \texttt{FROM} clause of the final SQL query, and is difficult because NLIDB users do not have a knowledge of the database schema or SQL and therefore cannot explicitly specify the intermediate tables and joins needed to construct a final SQL query.

Table~\ref{tab:nlidb} summarizes state-of-the-art systems and their strategy to handle each step of NLQ to SQL translation. The upper half lists pipeline-based systems, where each subproblem is explicitly handled, while the lower half are deep learning systems which implicitly tackle these challenges by the choice of input representation and network architecture. The {\em keyword mapping} task is split into the {\em Rel/Attr Mapping} and {\em Value Mapping} columns because some systems have independent procedures for handling each. Some common patterns emerge:

\begin{itemize}
\item For {\em keyword mapping}, the vast majority of systems make use of a lexical database such as WordNet~\cite{miller:wordnet} or a word embedding model~\cite{mikolov:word2vec,pennington2014glove}.
\item {\em Join path inference} is primarily handled via user interaction~\cite{li:nalir} or heuristics such as selecting the shortest join path~\cite{utama2018end} or hand-written repair rules~\cite{yaghmazadeh:sqlizer}.
\end{itemize}

\begin{figure}[t]
  \centering
  \includegraphics[width=0.85\columnwidth]{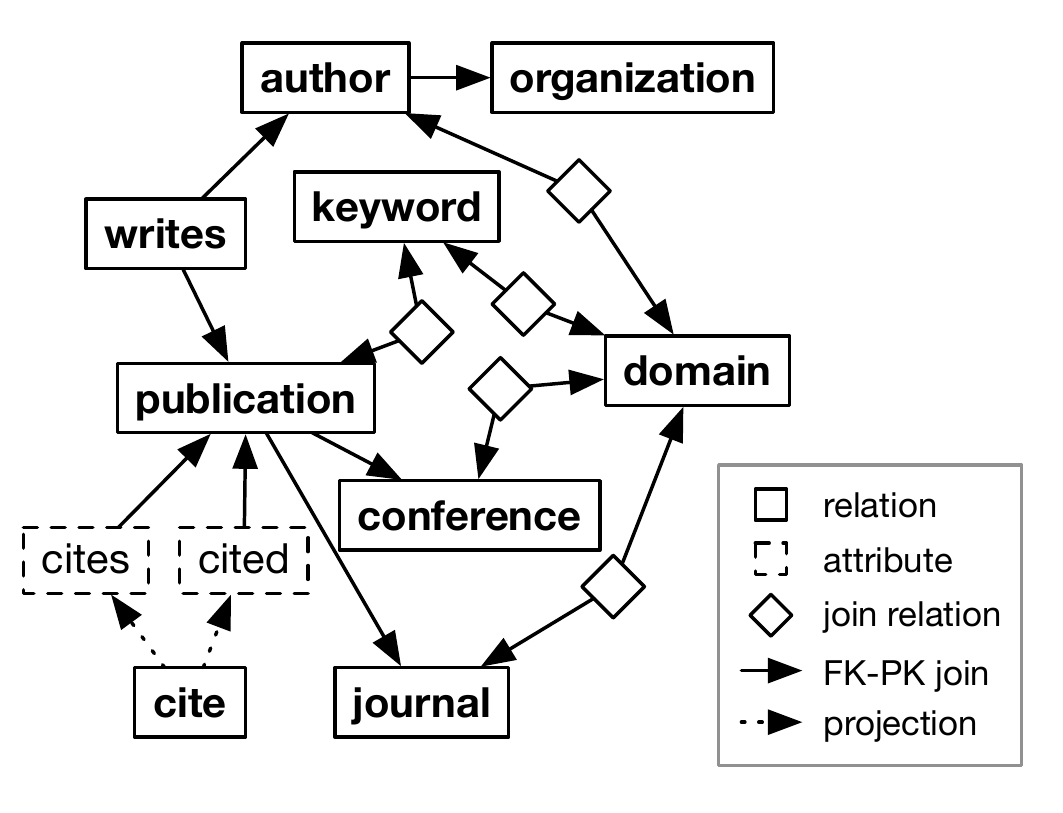}
  \vspace{-0.3cm}
  \caption{
  A simplified version of the Microsoft Academic Search database's schema graph.
  }
  \vspace{-0.5cm}
  \label{fig:mas}
\end{figure}

While each of these approaches works reasonably well, there is still significant room for improvement. For example:

\begin{example}
\label{ex:intro}
John issues an NLQ: ``Find papers in the Databases domain'' on an academic database (Figure~\ref{fig:mas}) using a pipeline NLIDB. John's intended SQL query is:

\begin{verbatim}
SELECT p.title
FROM publication p, publication_keyword pk
 keyword k, domain_keyword dk, domain d
WHERE d.name = `Databases'
 AND p.pid = pk.pid AND k.kid = pk.kid
 AND dk.kid = k.kid AND dk.did = d.did
\end{verbatim}

The NLIDB attempts \textbf{keyword mapping} by matching ``papers'' in the NLQ to either the relation \texttt{publication} or \texttt{journal}, and ``Databases'' to a value in the \texttt{domain} relation. It maps ``papers'' to \texttt{journal} because they have a high similarity score in the NLIDB's word embedding model. After this, the NLIDB performs \textbf{join path inference} by examining the schema graph and selects the shortest join path from \texttt{journal} to \texttt{domain} to form the (unintended) SQL query:

\begin{verbatim}
SELECT j.name
FROM journal j, domain_journal o, domain d
WHERE d.name = `Databases'
 AND j.jid = o.jid AND o.did = d.did
\end{verbatim}
\end{example}

The example demonstrates how error in keyword mapping can propagate through the pipeline to produce an incorrect SQL query. Even when the keyword mapping is correct, however, the join path inference remains as a challenge:

\begin{example}
In the keyword mapping process for John's NLQ, assume the NLIDB correctly matched ``papers'' to \texttt{publication}. The NLIDB examines the schema graph and its algorithm selects the shortest path from \texttt{publication} to \texttt{domain}. The returned SQL does not match John's intent:

\begin{verbatim}
SELECT p.title
FROM publication p, conference c,
 domain_conference dc, domain d
WHERE d.name = `Databases'
 AND p.pid = c.pid AND c.cid = dc.cid
 AND dc.did = d.did
\end{verbatim}
\end{example}

While there is always inherent ambiguity introduced in NLQs that even humans have difficulty interpreting, our goal is to improve the accuracy of {\em keyword mapping} and {\em join path inference} in NLIDBs to better match the user's intent.

Recent end-to-end deep learning systems~\cite{utama2018end,xu2017sqlnet,zhong2017seq2sql} show the great promise of learning from large volumes of NLQ-SQL pairs. However, manually creating labeled NLQ-SQL pairs is costly and time-consuming. Despite recent efforts to synthesize NLQ-SQL pairs~\cite{iyer2017learning,utama2018end} or derive them from user descriptions of SQL queries~\cite{brad2017dataset}, obtaining realistic labeled data remains an open research challenge. As a result, state-of-the-art deep learning systems~\cite{xu2017sqlnet,zhong2017seq2sql} have thus far only been tested on datasets of simple NLQs requiring no join.

\paragraph{Our Approach} While NLQ-SQL pairs are rarely available in large quantities for a given schema, large SQL query logs are more readily available given that NLIDBs are often built not for freshly instantiated databases, but for existing production databases~\cite{friendlydata,powerbi}. Although the SQL query log is not a typical supervised learning training set of input-output pairs, an output set of rich data like the SQL query log can still provide value in translation, akin to the way that one could infer much about what is being communicated and what should be spoken next even by listening to only one end of a phone conversation. Our approach is to use the information in the SQL query log of a database to select more likely keyword mappings and join paths for SQL translations of NLQs.

We propose a system \system, which augments existing pipeline-based NLIDBs such as \cite{li:nalir,popescu:precise,saha2016athena,yaghmazadeh:sqlizer} with SQL query log information. While it is also possible to augment end-to-end deep learning NLIDBs, this would require additional pre- or post-processing, and we leave it for future work.  Consider the user of \system\ with our running example:

\begin{example}
John issues the NLQ from Example~\ref{ex:intro} on a NLIDB augmented with \system. The NLIDB defers the \textbf{keyword mapping} to \system, which uses information in the SQL query log to determine \texttt{publication} as the most likely mapping. The NLIDB receives this information, performs any necessary processing, and then defers \textbf{join path inference} to \system\ by passing the mapped relations and attributes to it. \system\ takes the input and again uses the SQL query log to conclude that the most likely join path involves connecting \texttt{publication} to \texttt{domain} via the \texttt{keyword} relation. This join path is passed back to the NLIDB, which constructs the final SQL query matching John's intent.
\end{example}

\paragraph{Technical Challenges} Unlike traditional learning tasks where full input-output pairs (\ie\ NLQ-SQL pairs) are used to train a model, we use only output logs (\ie\ SQL queries). Consequently, the information in the SQL query log does not directly map to the translation task. Furthermore, even with large query logs, it is likely that most queries are not exact repeats of queries previously issued. Finally, our goal is to augment, rather than replace, NLIDBs. So we need \system\ to be able to assist multiple NLIDBs through a simple common interface. In short, the challenges are to
\begin{enumerate*}[label=(\arabic*)]
\item {\em selectively activate information in the SQL log} for NLQ-SQL translation,
\item {\em allow the generation of new SQL queries} not in the log, and
\item {\em gracefully integrate log information with existing techniques} in NLIDBs.
\end{enumerate*}

\paragraph{Contributions} 

Our main contributions are as follows:

\begin{itemize}
\item We propose the {\em query fragment} as an atomic building block for SQL, providing a fine-grained view of a SQL log to allow {\em selective activation of information in the log}. Query fragments can be mixed and matched to {\em allow the generation of new SQL queries} not yet observed in the query log.
\item We propose the {\em Query Fragment Graph} as a novel abstraction to enhance the accuracy of keyword mapping and join path inference in NLIDBs by modeling the co-occurrence of query fragments from a SQL query log, and {\em gracefully integrating this with existing techniques} to improve the accuracy of keyword mapping and join path inference in NLIDBs.
\item We introduce a prototype system \system, which augments existing NLIDBs without altering their internal architecture.
\item We demonstrate by an extensive evaluation on how \system\ can improve the top-1 accuracy of state-of-the-art NLIDBs by up to 138\% on our benchmarks.
\end{itemize}

\paragraph{Organization} We discuss related work (Section~\ref{sec:related}), then present the architecture of \system\ and formal problem definitions (Section~\ref{sec:overview}), before introducing the query fragment and Query Fragment Graph to model the SQL query log (Section~\ref{sec:log}). We then explain our algorithms for improving the accuracy of keyword mapping and join path inference by leveraging the Query Fragment Graph (Sections~\ref{sec:mapping}-\ref{sec:joinpath}). We present our experimental evaluation of NLIDBs augmented with \system\ (Section~\ref{sec:eval}), and conclude (Section~\ref{sec:concl}).

\section{Related Work}
\label{sec:related}

Our work builds upon multiple streams of prior work:

\paragraph{Natural language interfaces to databases (NLIDB)} Research on NLIDBs extends as far back to the sixties and seventies~\cite{androutsopoulos:nlidb}, when interfaces were focused on solutions tailored to a specific domain. Early approaches depended on grammars that were manually-specified~\cite{androutsopoulos:nlidb} or learned from database-specific training examples~\cite{ge2005statistical,tang2001using}, making it difficult to scale them across different database schemas.

Since then, advances in deep learning have inspired efforts to build an end-to-end deep learning framework to handle natural language queries~\cite{dong2016language,liang2016learning,yin2016neural}. The limiting factor for such systems is the need for a large set of NLQ to SQL pairs for each schema, and consequently some work focuses on the challenge of synthesizing and collecting NLQ-SQL pairs~\cite{brad2017dataset,iyer2017learning,utama2018end,yu2018spider} to be able to train these systems. The most recent deep learning-based end-to-end systems~\cite{utama2018end,xu2017sqlnet,zhong2017seq2sql} make use of the sequence-to-sequence architecture, and these systems can benefit from the enhancements \system\ provides to keyword mapping. On the other hand, the current scope of these systems is confined to single-table schemas which do not require joins, and therefore the join path inference techniques described in \system\ provide no benefit to these systems as yet.

An alternative approach has been to combine techniques from the natural language processing and database communities to construct pipeline-based NLIDBs. Such systems often utilize intermediate representations in the NLQ to SQL translation process, such as a parse tree~\cite{li:nalir}, query sketch~\cite{unger2012template,yaghmazadeh:sqlizer}, or an ontology~\cite{saha2016athena}. They also ensure reliability by doing at least one of the following: explicitly defining their semantic coverage~\cite{li:nalir,li2005nalix,popescu:precise}, allowing the user to correct ambiguities~\cite{li:nalir}, asking the user to provide a mapping from a database schema to an ontology~\cite{saha2016athena}, or by engaging in an automated query repair process~\cite{yaghmazadeh:sqlizer}. \system\ can enhance the performance of these NLIDBs by leveraging query logs as an additional data source to increase accuracy.

\paragraph{Keyword search} Keyword search interfaces~\cite{agrawal2002dbxplorer,bergamaschi2016combining,bhalotia2002keyword,blunschi2012soda,hristidis2002discover,tata2008sqak} emulate web search engines by allowing users to type in keywords to retrieve information. These keyword search interfaces often face the keyword mapping and join path inference problems that were described in our work, but \system\ is the first to make use of the SQL query log to address these issues.

\paragraph{Using query logs} Previous work used SQL query logs to autocomplete SQL queries~\cite{khoussainova2010snipsuggest}, proposing a similar abstraction to query fragments for a different purpose. QueRIE~\cite{eirinaki2014querie} and qunits~\cite{nandi2009qunits} organized the query log in a similar fashion to the Query Fragment Graph, but for the purposes of query recommendations and keyword queries, respectively.

\section{Overview}
\label{sec:overview}

\subsection{Preliminaries}
We first introduce some preliminary definitions.

The schema graph depicts the relations and their connections in a relational database:

\begin{definition}
A \textbf{schema graph} is a directed graph $G_s = (V,E,w)$ for a database $D$ with the following properties:
\begin{itemize}
\item $V$ consists of two types of vertices:
  \begin{itemize}
  \item Relation vertices $V_R \subseteq V$, each corresponding to a relation in $D$.
  \item Attribute vertices $V_\alpha \subset V$, each corresponding to an attribute in $D$.
  \end{itemize}
\item $E$ consists of two types of edges:
  \begin{itemize}
  \item Projection edges $E_\pi \subseteq E$, each extending from a given relation vertex to each of its corresponding attribute vertices.
  \item FK-PK join edges $E_{\bowtie} \subset E$, each extending from each foreign key attribute vertex to its corresponding primary key attribute vertex.
  \end{itemize}
\item $w: V \times V \rightarrow [0,1]$ is a function that assigns a weight to each pair of vertices which have an edge in $E$.
\end{itemize}
\end{definition}

A join path is a specific type of tree within the schema graph, which can be represented by a combination of relations and join conditions in a SQL query:

\begin{definition}
Given a schema graph $G_s$ and a bag of relations $B_R$, a \textbf{join path} $(V_j,E_j,V_t)$ is a tree of vertices $V_j \subset G_s$ and edges $E_j \subset G_s$ spanning all terminal vertices $V_t \subset G_s$, where each relation instance in $B_R$ is represented by a terminal vertex $v_R \in V_t$.
\end{definition}

\subsection{Definitions}

As we will discuss in detail in Sec.~\ref{sec:log}, a complete SQL query is too large and too specific a unit of data to be able to use it effectively to represent a SQL log.
Instead, we use {\em query fragments}, which are pieces of SQL queries:

\begin{definition}
A \textbf{query fragment} $\textit{c} = (\chi, \tau)$ is a pair of:
\begin{itemize}
\item $\chi$: a SQL expression or non-join condition predicate;
\item $\tau$: the context clause in which $\chi$ resides.
\end{itemize}
\end{definition}

For example, in the SQL query:

\begin{verbatim}
SELECT t.a FROM table1 t, table2 u
WHERE t.b = 15 AND t.id = u.id
\end{verbatim}

The query fragments are (\texttt{t.a}, \texttt{SELECT}), (\texttt{table1}, \texttt{FROM}), (\texttt{table2}, \texttt{FROM}), (\texttt{t.b = 15}, \texttt{WHERE}).

Keyword phrases in a NLQ are mapped to query fragments by NLIDBs to form {\em query fragment mappings}:

\begin{definition}
A \textbf{query fragment mapping} $\textit{m} = (s, c, \sigma)$ is a triple of a keyword $s$, a query fragment $c$, and a similarity score $\sigma$ between the keyword and query fragment.
\end{definition}

A selection of mappings for an NLQ form a {\em configuration}:

\begin{definition}
A \textbf{configuration} $\phi(S)$ of a set of keywords $S$ is a selection of exactly one query fragment mapping $(s_k, c_k,\sigma_k)$ for each keyword $s_k \in S$, where $c_k$ is a query fragment, and $\sigma_k$ is the associated similarity score for the keyword and fragment.
\end{definition}

\subsection{Problem Definitions}
We now present a formal definition for the {\em keyword mapping} and {\em join path inference} problems.

\subsubsection{Keyword Mapping}
\label{sec:keywordproblem}

The keyword mapping problem is described by the function:
$$\Phi = \textsc{MapKeywords}(D,S,M)$$

The input to the problem is a database $D$, a set of keywords representing an NLQ, $S = \{s_1, s_2, ..., s_n\}$, where each keyword $s_k \in S$ can be comprised of multiple words or tokens in natural language; and a set of metadata annotations, $M$, where each element $M_k = (\tau_k,\omega_k,\mathcal{F}_k,g_k)$ of $M$ includes parser metadata about $s_k$: the context $\tau_k$ of the query fragment that should be mapped to $s_k$, an optional predicate comparison operator $\omega_k$, an optional ordered list of aggregation functions $\mathcal{F}_k$, and a boolean $g_k$ which if true, indicates that the resulting mapping of $s_k$ should be grouped. The goal of the problem is to return a list of {\em configurations} $\Phi$ ordered by likelihood.

\subsubsection{Join Path Inference}

The join path inference problem is described by the function:
$$J = \textsc{InferJoins}(G_s, B_D)$$

The input is a schema graph $G_s$, a bag (\ie\ a multiset) of attributes and relations $B_D$ that are known to be part of the SQL query. The goal is to return a list of join paths $J$ on $G_s$ ranked from most to least likely.

\subsection{Architecture}

\begin{figure}[t]
  \centering
  \includegraphics[width=\columnwidth]{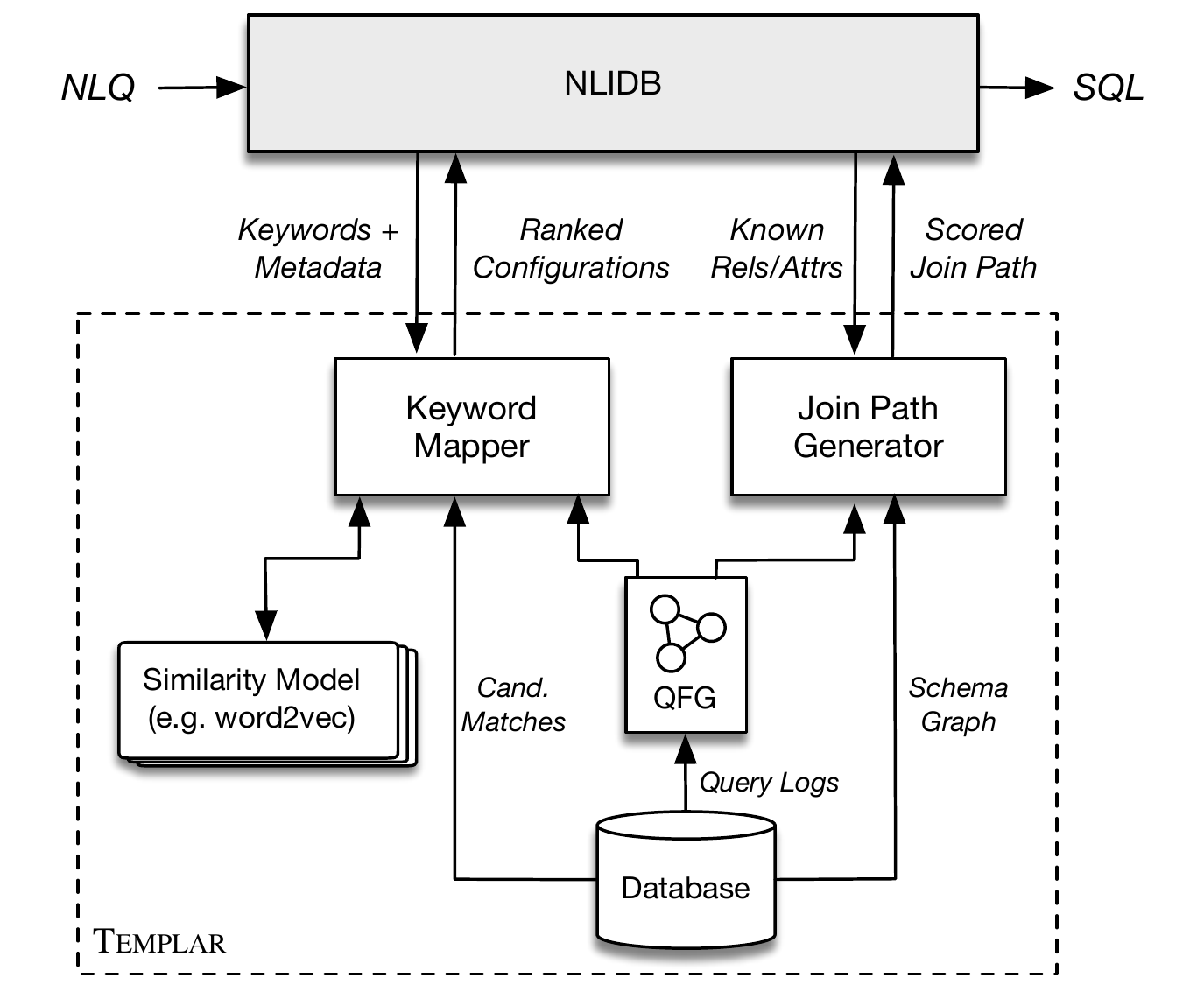}
  \vspace{-0.4cm}
  \caption{
  The overall architecture of an NLIDB augmented with \system.
  }
  \label{fig:arch}
  \vspace{-0.3cm}
\end{figure}

\system's architecture is shown in Figure~\ref{fig:arch}. It interfaces with the NLIDB it is augmenting on two fronts: one for keyword mapping, and the other for join path inference.

The \textbf{\em Keyword Mapper} carries out the execution of \textsc{MapKeywords}, and uses a word similarity model such as word2vec~\cite{mikolov:word2vec} or GloVe~\cite{pennington2014glove}, the query fragment graph (QFG) which stores the SQL query log information, and the database itself to retrieve candidate matches. The \textbf{\em Join Path Generator} executes \textsc{InferJoins}, and it utilizes the QFG and the schema graph of the database to infer join paths.

\subsection{NLIDB Prerequisites}
\label{sec:prereqs}

An NLIDB to which we can apply our approach is responsible for the following:

\begin{itemize}
\item It must be able to parse the NLQ into keywords, which may require recognition of multi-word entities. Each keyword should have associated metadata (query fragment type, predicate operator, aggregation functions, and presence of a group-by) for the keyword mapping problem.
\item It is responsible for constructing a SQL query given the keyword mappings and join paths provided by \system.
\end{itemize}

The categories of metadata we expect as input in {\sc MapKeywords} are all obtainable using existing parser technology~\cite{klein2003accurate,nivre2016universal,schuster2016enhanced} by existing NLIDBs~\cite{li:nalir,yaghmazadeh:sqlizer}.

Since the two main interface calls of keyword mapping and join path inference are independent of one another in our approach, we do not enforce any ordering of when and how these calls should be made within the NLIDB. However, in every currently known system in Table~\ref{tab:nlidb}, the keyword mapping step precedes the join path inference step.

The interface to pipeline-based NLIDBs such as \cite{li:nalir,popescu:precise,saha2016athena,yaghmazadeh:sqlizer} is transparent, as most already support the above requirements or can be easily modified to do so. Integrating \system\ into an end-to-end deep learning NLIDB is possible by integrating the information from the SQL query log into the input representation or by performing some pre-processing and/or post-processing, but we leave this for future work.

\subsection{Example Execution}

In this section, we describe an example execution of a generic pipeline-based NLIDB augmented with \system. Consider the architecture in Figure~\ref{fig:arch} and the following example NLQ from the Microsoft Academic Search (MAS) dataset~\cite{li:nalir} with schema in Figure~\ref{fig:mas}:

\begin{example}
\label{ex:overview}
Return the papers after 2000.
\end{example}

First, the NLIDB parses the NLQ to return the keywords which map to elements in the database and corresponding parser metadata. In Example~\ref{ex:overview}, the keywords emitted by the NLIDB would be {\em papers} and {\em after 2000}. NLIDBs have various techniques of producing the metadata, whether through semantic parsing~\cite{yaghmazadeh:sqlizer} or a designated lexicon of keywords~\cite{li:nalir}. The NLIDB in \cite{li:nalir} would return that {\em papers} is in the \texttt{SELECT} context because it is a direct child of the keyword {\em Return} in the parse tree, and {\em after 2000} would be in the \texttt{WHERE} context because {\em after} is a reserved keyword corresponding to the predicate comparison operator $>$ in the NLIDB's lexicon.

The keywords are passed to the \textbf{\em Keyword Mapper}, which maps each keyword to candidate query fragments using the keyword metadata and information about the database schema and contents. These candidate query fragment mappings are individually scored using a similarity model (such as word2vec~\cite{mikolov:word2vec}) and information from the {\em Query Fragment Graph (QFG)}. For Example~\ref{ex:overview}, the candidate mappings for {\em papers} includes (\texttt{journal.name}, \texttt{SELECT}) and (\texttt{publication.title}, \texttt{SELECT}), and {\em after 2000} is mapped to (\texttt{publication.year $>$ 2000}, \texttt{WHERE}).

A configuration is generated by selecting one candidate mapping per keyword. The top-$\kappa$ most likely candidate configurations are returned by the \textbf{\em Keyword Mapper}. Example~\ref{ex:overview} produces at least two candidate configurations, whose mapped query fragments, respectively, are:

\begin{itemize}
\item \texttt{[(journal.name, SELECT);\\ (publication.year $>$ 2000, WHERE)]}
\item \texttt{[(publication.title, SELECT);\\ (publication.year $>$ 2000, WHERE)]}
\end{itemize}

These configurations are then sent back to the NLIDB, which can augment the ranked configurations with other information such as domain-specific knowledge.

After processing the configurations, the NLIDB sends known relations for each candidate SQL translation to the \textbf{\em Join Path Generator}, which identifies the most likely join path and returns it along with an associated score.

For the schema graph shown in Figure~\ref{fig:mas} and continuing with Example~\ref{ex:overview}, this step will produce the join path \texttt{journal}-\texttt{publication} for our first configuration, and the single relation \texttt{publication} for the second.

Finally, it is the NLIDB's responsibility to construct the final SQL query and return it. Any post-processing, such as the hand-written repair rules in \cite{yaghmazadeh:sqlizer} or soliciting additional user interaction as in \cite{li:nalir} may also be performed at this point. For our running example, the final SQL queries returned by the NLIDB for each candidate configuration would be:

\begin{itemize}
\item \begin{verbatim}
SELECT j.name
FROM journal j, publication p
WHERE p.year > 2000 AND j.jid = p.jid
\end{verbatim}

\item \begin{verbatim}
SELECT title FROM publication
WHERE year > 2000
\end{verbatim}
\end{itemize}

\section{Query Log Model}
\label{sec:log}

\begin{figure}[t]
\subfloat[Example query log.]{%
  \includegraphics[clip,width=\columnwidth]{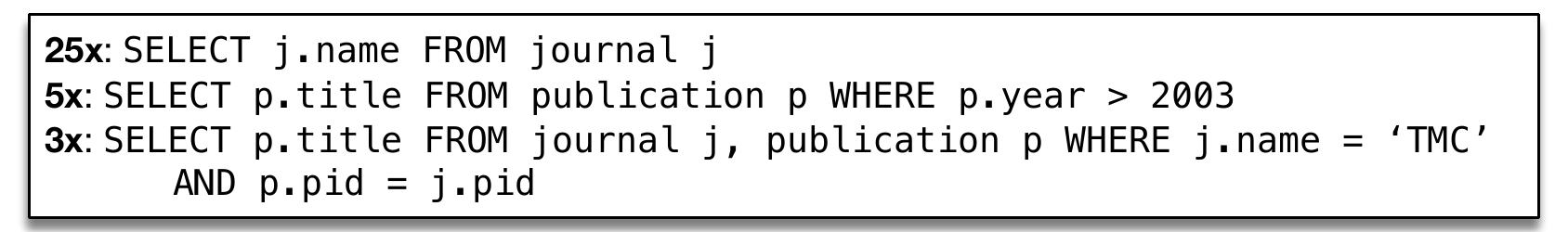}%
  \label{fig:log}
}

\subfloat[Query fragment occurrences.]{%
  \includegraphics[clip,width=\columnwidth]{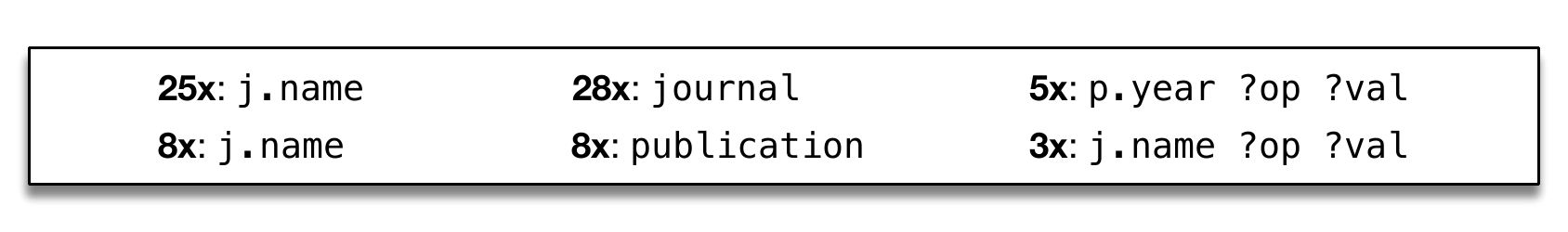}%
  \label{fig:freq}
}

\centering
\subfloat[Query fragment graph.]{%
  \includegraphics[clip,width=0.8\columnwidth]{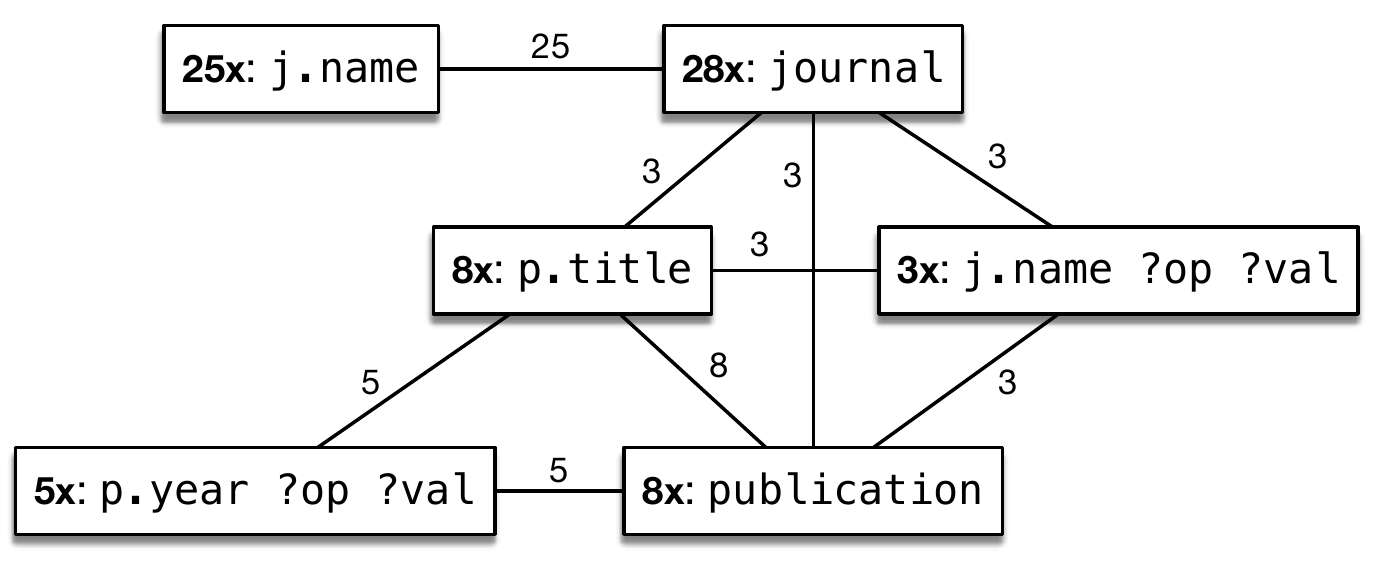}%
  \label{fig:qfg}
}

\caption{Storing query log information in the QFG.}
\vspace{-0.5cm}
\end{figure}

In this section, we explore how to model information in the SQL query log to aid in NLQ to SQL translation. Consider the SQL query log in Figure~\ref{fig:log} and the example task:

\begin{example}
\label{ex:log}
The task is:
\begin{itemize}
\item \textbf{NLQ}: Select all papers from TKDE after 1995.
\item \textbf{SQL}: \texttt{\em SELECT p.title\\ FROM journal j, publication p\\ WHERE j.name = `TKDE' AND\\ p.year > 1995 AND j.jid = p.jid}
\end{itemize}
\end{example}

First, we want {\em to generate queries not yet observed in the SQL log}---\ie\ not be constrained to only translate to queries already in the log. In Example~\ref{ex:log}, the NLQ has the keyword {\em papers} which might map to \texttt{publication.title} or \texttt{journal.name}. If we are limited to selecting existing SQL queries in the log to translate to, the NLQ could erroneously be translated to \texttt{SELECT j.name FROM journal j}.

To avoid this, we break down SQL queries into query fragments which can be mixed and matched to form new SQL queries, and count occurrences of each query fragment in the log as in Figure~\ref{fig:freq}.

Now, consider that we boost the scores of commonly-occurring query fragments in the SQL log. Unfortunately, there is still a high chance that ``papers'' will be mapped to \texttt{journal} because of its high frequency in the log.

Consequently, we want to {\em selectively activate information in the log} only when helpful for the NLQ at hand. The intuition is that the full NLQ provides context for each individual keyword, and this should be leveraged to illuminate what queries in the SQL log are relevant to the NLQ. In Example~\ref{ex:log}, the keywords are {\em papers}, {\em TKDE}, and {\em after 1995}. A human expert would that {\em TKDE} is referring to a journal and {\em after 1995} refers to a year, and can conclude that {\em papers} isn't referring to \texttt{journal.name} because the NLQ would be redundantly asking for ``all journals from a journal''.

Finally, we want to {\em maximize the semantic information in the SQL query log}. We observe a distinction between the more abstract semantic information and the specific value instances in a query. For Example~\ref{ex:log}, we can replace specific values in the NLQ with placeholders: {\em Select all papers from (journal) after (year)}, preserving the semantic structure while obscuring exact values. Similarly, we can put placeholders in the SQL:

\begin{verbatim}
SELECT p.title
FROM journal j, publication p
WHERE j.name ?op ?val AND p.year ?op ?val
 AND j.jid = p.jid
\end{verbatim}

Using such placeholders allows us to focus on the recurrence of semantic contexts without being distracted by specific values. Consequently, it allows us to make more extensive use of the data in the SQL query log as more query fragments in the log are likely to match any given keyword in a NLQ.

We implement three levels of {\em obscurity} for query fragments. The first level, {\em Full}, retains all values in the original query. The second, {\em NoConst}, replaces literal constants with a placeholder to convert a fragment \texttt{p.year > 2000} into \texttt{p.year > ?val}. Finally, we further obscure comparison operators in {\em NoConstOp} to make the fragment \texttt{p.year ?op ?val}.

\subsection{Query Fragment Graph}

While automated NLIDBs don't have the benefit of human logic, the SQL query log can play a similar role by using the full context of a NLQ to revise individual keyword mappings. Previous user queries in the log in Figure~\ref{fig:log} show that years are often queried in the context of \texttt{publication.title}, and similarly, when a specific journal name such as {\em TMC} is a predicate in a query, the user is often querying \texttt{publication.title}. This observation leads us to desire not only the occurrences of individual query fragments in the SQL log, but also the co-occurrences of query fragments---in other words, given the information in the SQL log, when one query fragment appears in a query, how likely is it that another query fragment is present in the query?

Given the intuition above, we introduce the \textbf{\em Query Fragment Graph (QFG)} as a data structure to store the information in a SQL query log.

\begin{definition}
A \textbf{query fragment graph} for database $D$ and SQL query log $L$ is a graph $G_f = (V_f, E_f, n_v, n_e)$ where:
\begin{enumerate}
\item each vertex $v \in V_f$ represents a query fragment in $L$;
\item each edge $e \in E_f$ exists if and only if two query fragments co-occur in $L$;
\item $n_v: V_f \rightarrow \mathbb{Z}_{\ge 0}$ is a function which maps $V_f$ to the number of occurrences in $L$ of the query fragment represented by each $v \in V_f$;
\item $n_e: V_f \times V_f \rightarrow \mathbb{Z}_{\ge 0}$ is a function which maps each pair of vertices to the co-occurrence frequency in $L$ of the two query fragments represented by the vertices.
\end{enumerate}
\end{definition}

In short, the QFG stores information on {\em query fragment occurrences} ($n_v$) in the log, as well as {\em co-occurrence relationships} ($n_e$) between each pair of query fragments.

\section{Keyword Mapping}
\label{sec:mapping}

In this section, we explain the keyword mapping procedure. While many techniques described here are already applied in existing work, we explain each step in detail to keep this work self-contained, and to clearly show how our novel approach of using SQL query log information comes into play.

Mapping keywords involves three steps:
\begin{enumerate*}[label=(\arabic*)]
\item retrieving candidate keyword to query fragment mappings,
\item scoring and retaining the top-$\kappa$ candidates, and
\item generating and scoring configurations.
\end{enumerate*}
Information from the query fragment graph is used in the final step to score configurations according to the evidence in the SQL query log.

We now describe our algorithm for the \textsc{MapKeywords} function, shown in Algorithm~\ref{alg:mapping}. We loop through all the keywords $s_k \in S$ with their corresponding metadata, then combine and rank them to form our output configurations.


\begin{algorithm}[t]
  \caption{Mapping Keywords}
  \label{alg:mapping}
  \begin{algorithmic}[1]
   \Function{MapKeywords}{$D$,$S$,$M$}
     \State $\mathcal{R} \gets \{\}$
     \For{$k \gets 1,\ldots,|S|$}
       \State $(\tau_k,\omega_k,\mathcal{F}_k,g_k) \gets M_k$
       \State $C_k \gets \textsc{KeywordCands}(D,s_k,\tau_k,\omega_k,\mathcal{F}_k,g_k)$
       \State $R_k \gets \textsc{ScoreAndPrune}(s_k,C_k,\kappa)$
       \State $\mathcal{R}.\textit{add}(R_k)$
     \EndFor
     \State $\Phi = \textit{genAndScoreConfigs}(\mathcal{R})$ \label{algline:scoreconfigs}
     \State \Return $\Phi$
    \EndFunction
  \end{algorithmic}
\end{algorithm}

\subsection{Retrieving Candidate Mappings}


\begin{algorithm}[t]
  \caption{Retrieve Candidate Keyword Mappings}
  \label{alg:kwcands}
  \begin{algorithmic}[1]
    \Function{KeywordCands}{$D$,$s$,$\tau$,$\omega$,$\mathcal{F}$,$g$}
      \State $C \gets \{\}$
      \If{$\textit{containsNumber}(s)$} \label{algline:num}
        \State $s_\textit{num} \gets \textit{extractNumber}(s)$
        \State $\beta \gets \textit{findNumericAttrs}(s_\textit{num}, \omega)$ \label{algline:findnum}
        \For{$b \in \beta$}
          \State $C.\textit{add}((\textsf{Pred}(b,\omega,s_\textit{num}), \texttt{WHERE}))$
        \EndFor
      \Else
        \If{$\tau = \texttt{FROM}$}
          \For{$r \in \textit{getRelations}(D)$}
            \State $C.\textit{add}((r, \tau))$
          \EndFor
        \ElsIf{$\tau = \texttt{SELECT}$}
          \For{$\alpha \in \textit{getAttributes}(D)$}
            \State $C.\textit{add}((\textsf{Attr}(\alpha,\mathcal{F},g), \tau))$
          \EndFor
        \Else
          \For{$t \in \textit{findTextAttrs}(s)$} \label{algline:findtext}
            \State $C.\textit{add}((\textsf{Pred}(t,=,s), \texttt{WHERE}))$
          \EndFor
        \EndIf
      \EndIf
      \State \Return $C$
   \EndFunction
  \end{algorithmic}
\end{algorithm}

The function {\sc KeywordCands} in Algorithm~\ref{alg:kwcands} maps a keyword $s$, along with its associated metadata $(\tau,\omega,\mathcal{F},g)$, to its candidate mappings $C$ by querying the database $D$.

First, we evaluate whether $s$ contains a number (Line~\ref{algline:num}), such as in the keyword {\em after 2000}. If so, we return all numeric attributes in the database that match a predicate formed by the number extracted from $s$ with the operator $\omega$ for $s$ (Line~\ref{algline:findnum}). For the keyword {\em after 2000}, we return all attributes containing at least one value that satisfies the predicate \texttt{?attr > 2000}. Predicates are constructed from matching attributes and added to the candidate set $C$.

If $s$ does not contain a number, we have three different cases. In the first two cases, where the context $\tau$ of the query fragment is \texttt{FROM} or \texttt{SELECT}, we simply add either all the relations or all the attributes (along with relevant metadata) of $D$ to the candidate set $C$.

For the final case covering all other structures, we first run a full-text search with every Porter-stemmed~\cite{van1980new} whitespace-separated token in $s$ to retrieve all matching text attributes $T$ in $D$ ({\em findTextAttrs} in Line~\ref{algline:findtext}). For example, for the keyword {\em restaurant businesses}, the stemming procedure would result in the tokens {\em restaur busi}, and we run the following SQL query, replacing \texttt{?attr} with each text attribute in $D$:

\begin{verbatim}
SELECT DISTINCT(?attr) FROM ?rel
WHERE MATCH(?attr)
 AGAINST (`+restaur* +busi*'
  IN BOOLEAN MODE)
\end{verbatim}

If any of the stemmed tokens from $s$ exactly match the stemmed attribute or relation names of a candidate query fragment, we remove them so as not to unnecessarily constrain our search. For example, if the keyword is {\em movie Saving Private Ryan} and a candidate query fragment mapping is an attribute from the \texttt{movie} relation, we remove the token {\em movie} from our full-text search query when searching on that attribute. For each matching text attribute, we then construct a predicate for the \texttt{WHERE} context.

\subsection{Scoring and Pruning}


\begin{algorithm}[t]
  \caption{Score and Prune Keyword Mappings}
  \label{alg:prune}
  \begin{algorithmic}[1]
   \Function{ScoreAndPrune}{$s$,$C$,$\kappa$}
      \State $R \gets \{\}$
      \For{$c \in C$}
        \If{$\textit{containsNumber}(s)$}
          \State $s_\textit{num} \gets \textit{extractNumber}(s)$
          \State $s_\textit{text} \gets s - s_\textit{num}$
          \State $\sigma \gets \textit{sim}_\textit{num}(s_\textit{text}, c)$
        \Else
          \State $\sigma \gets \textit{sim}_\textit{text}(s, c)$ \label{algline:sim}
        \EndIf
        \State $R.\textit{add}((s, c, \sigma))$
      \EndFor
      \State $\textbf{sort}\ R\textnormal{ by descending }\sigma$
      \State \Return $\textsc{Prune}(R,\kappa)$ \label{algline:prune}
    \EndFunction
  \end{algorithmic}
\end{algorithm}

Our next step is to retain only the top-$\kappa$ most likely mappings from $C$ with the function \textsc{ScoreAndPrune}.

We calculate a score $\sigma$ for each keyword mapping in the range $[0,1]$. For comparing keywords with purely text tokens against relation and attribute names and text predicates, we can use a similarity function $\textit{sim}_\textit{text}$ (Line~\ref{algline:sim}) through a word embedding model such as word2vec~\cite{mikolov:word2vec} or GloVe~\cite{pennington2014glove}. For keywords including numeric tokens, we execute (\ie\ $\textit{exec}(c)$) the candidate predicate on the database, then evaluate the similarity of only the text tokens if the predicate returns a non-empty set, and return a small $\epsilon$ value otherwise:
\[
    \textit{sim}_\textit{num}(s_\textit{text}, c) =
\begin{cases}
    \textit{sim}_\textit{text}(s_\textit{text},c),&\textnormal{if }\textit{exec}(c) \nrightarrow \varnothing \\
    \epsilon,&\textnormal{otherwise}\\
\end{cases}
\]

$\sigma$ is then combined into a tuple with the original keyword $s$ and candidate mapping $c$ and added to the result set $R$, which is finally sorted by descending $\sigma$ score. We then prune $R$ to prevent a combinatorial explosion when generating configurations, using the following \textsc{Prune} procedure (Line \ref{algline:prune}):

\begin{itemize}
\item If there are any candidates in $R$ that are exact matches ($\sigma \geq 1 - \epsilon$ for a small $\epsilon$), we prune away all remaining non-exact candidates.
\item Otherwise, we prune $R$ to the top-$\kappa$ results, including any results that have a non-zero $\sigma$ value that is equal to the $\sigma$ of the candidate at the $\kappa$-th place.
\end{itemize}

\subsection{Ranking Configurations}
\label{sec:configs}

At this point, we have a set of candidate mappings for each keyword $s_k \in S$. We combine and score them (Line~\ref{algline:scoreconfigs} of Algorithm~\ref{alg:mapping}) to form candidate configurations for $S$. We first describe a standard way of scoring configurations, then show how we can apply the SQL query log to improve scoring.

\subsubsection{Word Similarity-Based Score}
\label{sec:nlranking}

A \naive\ scoring function for configurations selects the best mapping for each keyword independently. We can take the geometric mean of the scores of all mappings to accomplish this:
$$\textit{Score}_{\sigma}(\phi) = [\prod_{(s_k,c_k,\sigma_k) \in \phi} \sigma_k]^{\frac{1}{|\phi|}}$$

We prefer the geometric mean over the arithmetic mean, as in \cite{yaghmazadeh:sqlizer}, to mitigate the impact of the variation in ranges of values for each keyword's candidate mapping scores.

\subsubsection{Query Log-Driven Score}
\label{sec:logranking}

Since we have the query log information available to us via the Query Fragment Graph, we leverage this information to derive an improved scoring function contextualized for our specific database schema.

While word similarity-based scoring considers each mapping independently, we now consider the {\em collective score} of each configuration of mappings. Previous work such as \cite{li:nalir} attempts a collective scoring approach based on {\em mutual relevance} which considers the proximity of keywords in the natural language dependency tree in relation to the edge weights connecting the candidate query fragments within the schema graph. Unfortunately, these schema graph edge weights are assigned manually without justification.

In contrast, the intuition behind our collective scoring mechanism is to give a higher score to configurations containing query fragments that frequently co-occur in queries in the SQL query log. Instead of relying on the system administrator's ability to preset the schema graph edge weights to match an anticipated workload, we derive our scoring directly from previous users' queries in the SQL query log.

To accomplish this, we calculate a metric for the co-occurrence of pairs of query fragments in the QFG, then aggregate this metric, along with the previously-computed similarity scores, over all query fragments in the configuration to derive a final score. We use the Dice similarity coefficient~\cite{gomaa2013survey} to reflect the co-occurrence of two query fragments $c_1$ and $c_2$ in the QFG, defined as follows:
$$\textit{Dice}(c_1,c_2) = \frac{2 \times n_e(c_1,c_2)}{n_v(c_1) + n_v(c_2)}$$

We accumulate {\em Dice} for every pair of non-relation (\ie\ not in the \texttt{FROM} context) fragments $(c_1,c_2) \in \phi_{\tau \ne \texttt{FROM}} \times \phi_{\tau \ne \texttt{FROM}}$:
$$\textit{Score}_\textit{QFG}(\phi) = [\prod_{(c_1,c_2) \in \phi^2_{\tau \ne \texttt{FROM}}} \textit{Dice}(c_1,c_2)]^{\frac{1}{|\phi|}}$$

The query fragments in the \texttt{FROM} context are excluded because involving relations can add information skewing the aggregate score---\eg\ if \texttt{journal.name} is in a SQL query, then the relation \texttt{journal} is required to be by the rules of SQL, adding unnecessary redundancy to the aggregated Dice score. In addition, relations in the \texttt{FROM} clause are explicitly handled by our join path inference procedure, so we defer the evaluation of these query fragments for later.

Finally, we perform a linear combination (governed by a parameter $\lambda \in [0,1]$) of $\textit{Score}_\sigma$ and the query log-driven score $\textit{Score}_\textit{QFG}$ to produce a final configuration score:
$$\textit{Score}(\phi) = \lambda \textit{Score}_\sigma(\phi) + (1 - \lambda) \textit{Score}_\textit{QFG}(\phi)$$

We can also replace this means of combining evidence from multiple sources with other approaches, such as the Dempster Shafer Theory in \cite{bergamaschi2016combining}. We opt for a linear combination due to its simplicity and because it works sufficiently well in practice.

All configurations are now scored using $\textit{Score}(\phi)$, ranked by descending score, and returned by \textsc{MapKeywords}.

\section{Join Path Inference}
\label{sec:joinpath}

In this section, we describe how we generate join paths for a set of attributes and relations selected to be part of the final SQL query by the keyword mapping procedure, and show how we use the SQL query log to improve this process.

\begin{example}
\label{ex:joinpath}
Consider that the NLIDB selected the following query fragments to be part of a SQL query of the schema given in Figure~\ref{fig:mas}:
\begin{itemize}
\item (\texttt{publication.title}, \texttt{SELECT})
\item (\texttt{domain.name = `Databases'}, \texttt{WHERE})
\end{itemize}

\textsc{InferJoins} should output the desired join path:

\texttt{publication}--\texttt{publication\char`_keyword}--\texttt{keyword}--\texttt{domain\char`_keyword}--\texttt{domain}.
\end{example}

\subsection{Generating Join Paths}

The process of generating the set of optimal join paths from a set of known relations $B_R$ and a schema graph $G_s$ has previously been modeled as the Steiner tree problem~\cite{kou1981fast}, where the goal is to find a tree on a graph that spans a given set of vertices with minimal total weight on its edges.

The expected input to the \textbf{\em Join Path Generator} is a bag of the attributes and relations $B_D$ already known to be in the desired SQL translation. $B_D$ can be converted to the bag of known relations $B_R$ simply by replacing each attribute with its parent relation in $G_s$.

We use a known algorithm~\cite{kou1981fast} for solving Steiner trees to find the set of optimal join paths for any given configuration. These optimal join paths, however, change depending on how weights are assigned to edges in the schema graph. We outline two ways to do this, first without information from the query log, and then adding in query log information.

\subsubsection{Default Edge Weights}

The default weight function $w$ for edges in the schema graph is to assign every edge a weight of 1. If we solve the Steiner tree problem with this weight function, we are essentially finding join paths with the minimal number of join edges that span all the known relations.

For Example~\ref{ex:joinpath}, this approach will produce the shortest join path between \texttt{publication} and \texttt{domain}, which is either:
\begin{itemize}
\item \texttt{publication}--\texttt{conference}--\texttt{domain\char`_conference}--\texttt{domain}
\item \texttt{publication}--\texttt{journal}--\texttt{domain\char`_journal}--\texttt{domain}
\end{itemize}
Neither of these join paths are the one desired by the user.

\subsubsection{Query Log-Driven Edge Weights}
\label{sec:logedge}

We look to the query log to provide some grounding for generating join paths. In contrast to previous work which depends on the system administrator to set schema graph edge weights~\cite{li:nalir}, on hand-written repair rules~\cite{yaghmazadeh:sqlizer}, or a predefined ontology~\cite{saha2016athena}, query log information is driven by actual user queries executed on the system. Query log information allows us to prefer commonly queried join paths, even if they are longer, and also mitigates the number of situations where there are identical scores given to equal-length join paths.

We leverage the co-occurrence values of relations in the QFG to adjust the weights on the schema graph. Given any two vertices $(v_1,v_2) \in G_s$, and the function $\textit{q}: V \rightarrow V_{QF}$ which maps a vertex in the schema graph $G_s$ to its corresponding vertex in the QFG, the new weight function is:
\[
    w_\textit{L}(v_1,v_2)= 
\begin{cases}
    1 - \textit{Dice}(q(v_1),q(v_2))& \text{if } v_1 \in V_R \land v_2 \in V_R\\
    1,& \text{otherwise}
\end{cases}
\]

This query log-based weight function $w_\textit{L}$ returns a lower value for join edges that frequently occur in the query log.

\subsection{Scoring Join Paths}

The final score for any join path $j$ we return is derived from the weights of the edges within the join path:
$$\textit{Score}_j(j) = \frac{1}{|E_j|^2} \sum_{(v_1,v_2) \in E_j} w(v_1,v_2)$$

We divide by $|E_j|^2$ to normalize the score in a [0,1] range and also to prefer simpler join paths over more complex ones. This is based on the observations regarding {\em semantic relevance}~\cite{bergamaschi2016combining,li:nalir} that the closer two relations are in the schema, the likelier it is that they are semantically related.

\subsection{Self-Joins}

\begin{figure}[t]
\centering
\subfloat[Before fork.]{
  \includegraphics[clip,width=0.32\columnwidth]{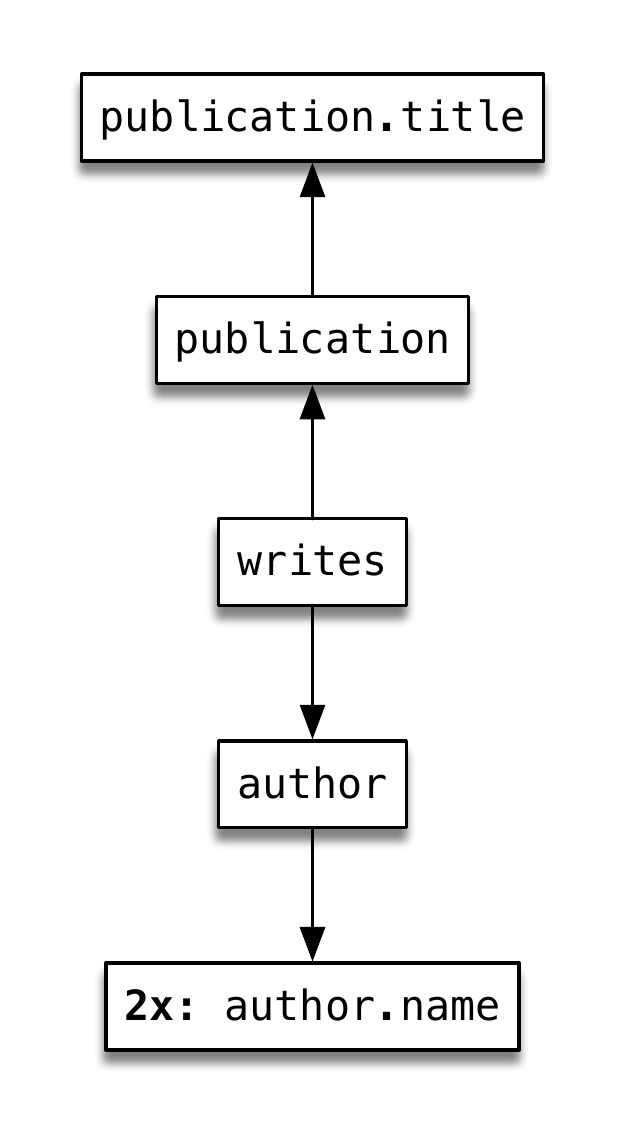}
  \label{fig:prefork}
}\qquad%
\subfloat[After fork.]{
  \includegraphics[clip,width=0.51\columnwidth]{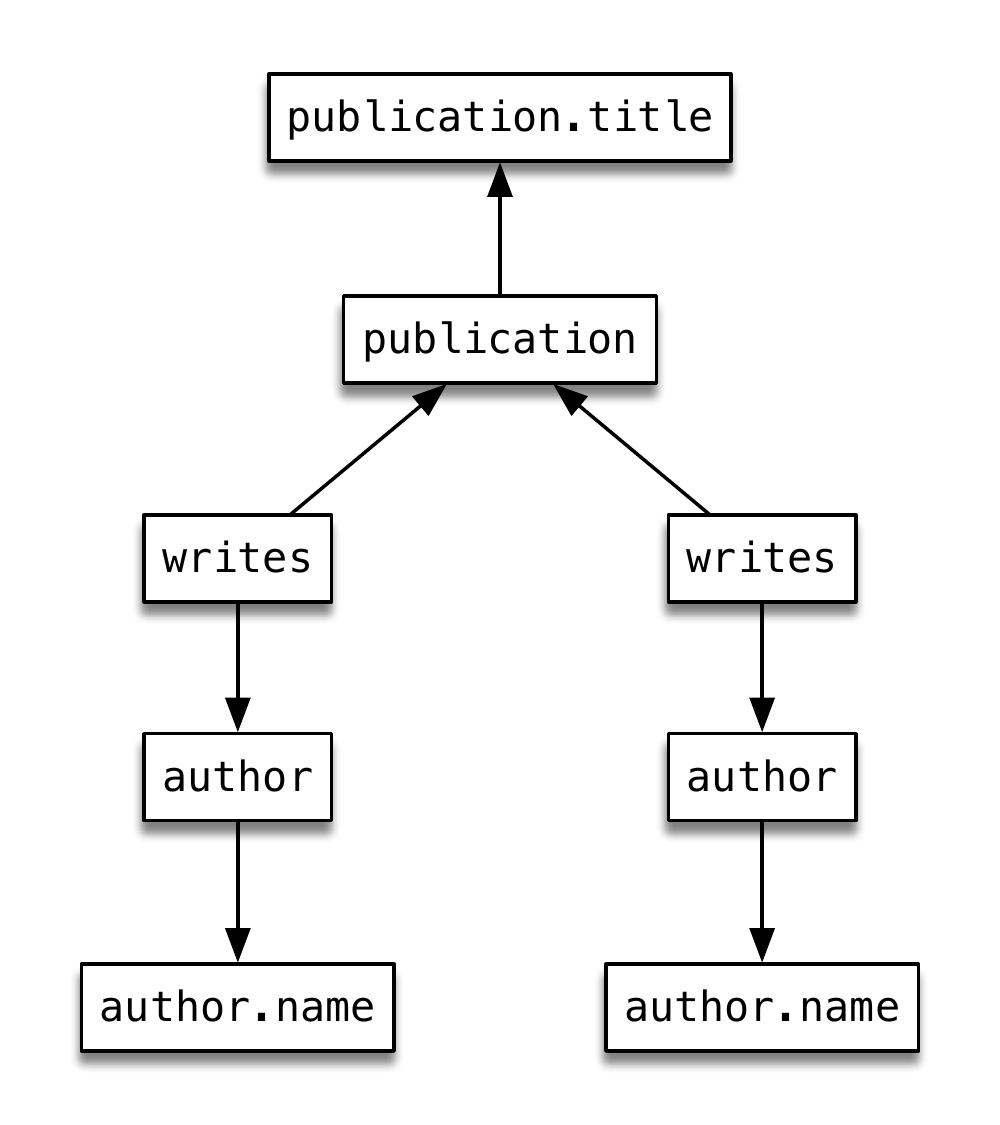}
  \label{fig:postfork}
}
\caption{A simplified overview of a schema graph fork for self-joins.}
\label{fig:fork}
\vspace{-0.2cm}
\end{figure}

A challenge arises during join path inference when an attribute is included multiple times in the bag $B_D$. We present a novel approach to handling such situations to still produce valid results from the Steiner tree algorithm.

Due to the peculiarities of SQL, these situations require that our resulting join path include multiple instances of the same relation, resulting in a {\em self-join}. For example:

\begin{example}
In an NLQ for the academic database, ``Find papers written by both John and Jane'', ``John'' and ``Jane'' both refer to attribute \texttt{\em author.name}. The correct SQL output for this NLQ is:

\begin{verbatim}
SELECT p.title
FROM author a1, author a2, publication p,
  writes w1, writes w2
WHERE a1.name = 'John' AND a2.name = 'Jane'
  AND a1.aid = w1.aid AND a2.aid = w2.aid
  AND p.pid = w1.pid AND p.pid = w2.pid
\end{verbatim}
\end{example}


\begin{algorithm}[t]
  \caption{Forking Schema Graph for Self-Joins}
  \label{alg:fork}
  \begin{algorithmic}[1]
    \Function{Fork}{$G_s$,$v$}
      \State $\textit{stack}_\textit{old} \gets \textbf{new}\ \textsf{Stack}()$
      \State $\textit{stack}_\textit{new} \gets \textbf{new}\ \textsf{Stack}()$
      \State $\textit{stack}_\textit{old}.\textit{push}(v)$
      \State $\textit{stack}_\textit{new}.\textit{push}(G_s.\textit{clone}(v))$ \label{algline:clone}
      \State $\textit{visited} \gets \{\}$
      \While{$\textit{stack}_\textit{old} \ne \varnothing$}
        \State $v_\textit{old} \gets \textit{stack}_\textit{old}.\textit{pop}()$
        \State $v_\textit{new} \gets \textit{stack}_\textit{new}.\textit{pop}()$
        \State $\textit{visited} \gets \textit{visited} \cup v_\textit{old}$
        \ForAll{$v_\textit{conn}\ \textnormal{connected to } v_\textit{old}$}
          \IIf{$v_\textit{conn} \in \textit{visited}$} $\textbf{continue}$
          \If{$(v_\textit{old}, v_\textit{conn}) \in E_{\bowtie}\textnormal{ of }G_s$} \label{algline:termfork}
            \State $\textbf{add edge}\ (v_\textit{new},v_\textit{conn}) \textnormal{ to } G_s$
          \Else
            \State $v_\textit{cloned} \gets G_s.\textit{clone}(v_\textit{conn})$ \label{algline:startfork}
            \State $\textit{dir} \gets \textnormal{direction of }(v_\textit{old},v_\textit{conn})$ \Comment{$\leftarrow$ or $\rightarrow$}
            \State $\textbf{add edge}\ (v_\textit{new},v_\textit{cloned},\textit{dir}) \textnormal{ to } G_s$
            \State $\textit{stack}_\textit{old}.\textit{push}(v_\textit{conn})$
            \State $\textit{stack}_\textit{new}.\textit{push}(v_\textit{cloned})$ \label{algline:endfork}
          \EndIf
        \EndFor
      \EndWhile
    \EndFunction
  \end{algorithmic}
\end{algorithm}

For these situations, we ``fork'' the schema graph, as shown (with some attribute vertices and edges removed for simplicity) in Figure~\ref{fig:fork}, in order to account for the necessary vertices for a join path containing a self-join.

Algorithm~\ref{alg:fork} describes the process of forking the schema graph $G_s$ in more detail, given an attribute vertex $v$ that has been referenced multiple times. Two mirrored stacks $v_\textit{old}$ and $v_\textit{new}$ are used to track progress for the original graph and the new fork of the graph, respectively. We first clone the attribute vertex $v$ and add it to $G_s$ (Line \ref{algline:clone}). We repeatedly pop the top of each stack, and find all vertices $v_\textit{conn}$ that are connected to the current existing vertex $v_\textit{old}$. We clone each $v_\textit{conn}$ and the edge connecting it to $v_\textit{old}$, then add both to the schema graph and continue traversal (Lines \ref{algline:startfork}-\ref{algline:endfork}). We terminate the forking process when we reach a FK-PK join edge in the direction {\em from} $v_\textit{old}$ {\em to} $v_\textit{conn}$ (Line \ref{algline:termfork}). For $d$ duplicate references to an attribute vertex $v$, \textsc{Fork} is executed $(d-1)$ times to create a fork for each duplicate reference.


\section{Evaluation}
\label{sec:eval}

We performed an experimental evaluation of our system, \system, to test whether we can use the SQL log to improve the accuracy of NLQ to SQL translation.

\subsection{Experimental Setting}

\subsubsection{Machine Specifications}
All our evaluations were performed on a computer with an 3.1 GHz Intel Core i7 processor and 16 GB RAM, running Mac OS Sierra.

\subsubsection{Compared Systems}
\label{sec:compared}

We enhanced two different NLIDB systems, NaLIR~\cite{li:nalir} and Pipeline, with \system, and executed them on our benchmarks. The augmented versions are denoted NaLIR+ and Pipeline+ respectively.

The first system we augmented is NaLIR~\cite{li:nalir}, a state-of-the-art pipeline-based NLIDB. We evaluated the system in its non-interactive setting because its application of user interaction is orthogonal to our approach.

We contacted authors of a few other existing NLIDBs but were not granted access to their systems. As a result, we built an NLIDB named Pipeline, which is an implementation of the keyword mapping and join path inference steps from the state-of-the-art approach in \cite{yaghmazadeh:sqlizer}, excluding the hand-written repair rules. Pipeline was implemented using word2vec~\cite{mikolov:word2vec} for keyword mapping, with the default Google News corpus for calculating word similarity. While the default similarity value produced from word2vec is a cosine similarity value in the range [-1, 1], Pipeline normalizes these values to fall in the range [0, 1]. Pipeline also always selects the minimum-length join paths for join path inference. Our implementation of Pipeline was written in Java. We used MySQL Server 5.7.18 as our relational database.

\subsubsection{Assumptions}

We assume \system\ is applied in a setting where {\em queries in the SQL query log are representative of the SQL queries issued by users via natural language}. While this assumption does not hold true for all databases, we believe \system\ is applicable for databases which already implement user-friendly interfaces such as forms or keyword search where the pattern of users' information need is likely to be similar to that of natural language interfaces.

\subsubsection{Dataset}

\begin{table}[t]
  \centering
  \caption{Statistics of each benchmark dataset.}
  \small
  \begin{tabular}{cccccc}
    \toprule
      \textbf{Dataset} & \textbf{Size} & \textbf{Rels} & \textbf{Attrs} & \textbf{FK-PK} & \textbf{Queries} \\
    \midrule
      MAS & 3.2 GB & 17 & 53 & 19 & 194 \\
      Yelp & 2.0 GB & 7 & 38 & 7 & 127 \\
      IMDB & 1.3 GB & 16 & 65 & 20 & 128 \\
    \bottomrule
  \end{tabular}
  \vspace{-0.3cm}
  \label{tab:benchmarks}
\end{table}

We tested each system by evaluating its ability to translate NLQs accurately to SQL on three benchmarks: the Microsoft Academic Search (MAS) database used in \cite{li:nalir}, and two additional databases from \cite{yaghmazadeh:sqlizer} regarding business reviews from Yelp and movie information from IMDB. Table~\ref{tab:benchmarks} provides some statistics on each of these benchmark datasets.

We annotated each NLQ with SQL translation by hand as the original benchmarks did not include the translated SQL queries. We removed 2 queries from MAS, 1 query from Yelp, and 3 queries from IMDB because they were overly complex (\ie\ contained correlated nested subqueries) or ambiguous, even for a human annotator.

We used a cross-validation method to ensure that the test queries were not part of the SQL query log used to perform the NLQ to SQL translation. Specifically, we randomly split the full dataset into 4 equally-sized folds, and performed 4 trials (one for each fold), where in each trial, the training set is comprised of 3 of the folds and the test set was the remaining fold held out of the training process. Our displayed results for all experiments are aggregated from the 4 trials.

For Pipeline and Pipeline+, we hand-parsed each NLQ into keywords and metadata to avoid any parser-related performance issues outside the scope of our work, while we passed the whole NLQ as input to NaLIR and NaLIR+ to make use of the authors' original system. For fairer comparison, we rewrote some NLQs with {\em wh-words} such as {\em who}, {\em what}, \etc\ to enable NaLIR/NaLIR+'s parser to process them correctly.


\begin{table}[t]
  \centering
  \small
  \caption{Keyword mapping (KW) and full query (FQ) results.}
  \begin{tabular}{clcc}
    \toprule
      \textbf{Dataset} &
      \textbf{System} &
      \textbf{KW (\%)} &
      \textbf{FQ (\%)} \\
    \midrule
      \multirow{4}{*}{MAS} &
      NaLIR & 43.3 & 33.0 \\
      & NaLIR+ & 45.4 & 40.2 \\
      & Pipeline & 39.7 & 32.0 \\
      & Pipeline+ & \textbf{77.8} & \textbf{76.3} \\
    \midrule
      \multirow{4}{*}{Yelp} &
      NaLIR & 52.8 & 47.2 \\
      & NaLIR+ & 59.8 & 52.8 \\
      & Pipeline & 56.7 & 54.3 \\
      & Pipeline+ & \textbf{85.0} & \textbf{85.0} \\
    \midrule
      \multirow{4}{*}{IMDB} &
      NaLIR & 40.6 & 38.3 \\
      & NaLIR+ & 57.8 & 50.0 \\
      & Pipeline & 32.0 & 27.3 \\
      & Pipeline+ & \textbf{67.2} & \textbf{64.8} \\
    \bottomrule
  \end{tabular}
  \label{tab:breakdown}
  \vspace{-0.5cm}
\end{table}


\begin{table}[t]
  \centering
  \small
  \caption{Improvement from activating log-based joins in Pipeline+.}
  \begin{tabular}{ccc}
    \toprule
      \textbf{Dataset} &
      \textbf{LogJoin} &
      \textbf{FQ (\%)} \\
    \midrule
      \multirow{2}{*}{MAS} &
      N & 68.6 \\
      & Y & \textbf{76.3} \\
    \midrule
      \multirow{2}{*}{Yelp} &
      N & 68.5 \\
      & Y & \textbf{85.0} \\
    \midrule
      \multirow{2}{*}{IMDB} &
      N & 60.9 \\
      & Y & \textbf{64.8} \\
    \bottomrule
  \end{tabular}
  \label{tab:logjoin}
  \vspace{-0.5cm}
\end{table}

\subsubsection{Evaluation Metrics}

We measured accuracy by checking the top-ranked SQL query returned by each system by hand. For Pipeline and Pipeline+, since it was possible to return multiple queries tied for the top spot, we considered the resulting queries incorrect if there were any tie for first place.

\subsection{Effectiveness of \system\ Augmentation}

In Table~\ref{tab:breakdown}, we present the overall performance of each system. Pipeline+ and NaLIR+ were both executed with obscurity \textit{NoConstOp}, $\kappa = 5$, and $\lambda = 0.8$. While all obscurity levels, including \textit{Full} and \textit{NoConst}, consistently improved on the baseline systems, we only show results for the best-performing obscurity level \textit{NoConstOp} for space reasons.

\subsubsection{Full Query}

The full query (FQ) was considered correct if the NLIDB ultimately produced the correct SQL query. Pipeline+ achieves 76.3\% accuracy on MAS, 85.0\% accuracy on Yelp, and 64.8\% accuracy on IMDB. Compared to the vanilla Pipeline system, this was a 138\%, 57\%, and 137\% increase in accuracy, respectively. NaLIR+ improved on NaLIR by more modest margins, with a 22\% increase for MAS, 12\% for Yelp, and 31\% for IMDB.

\subsubsection{Keyword Mapping}

For keyword mapping (KW), we considered the mapping correct if and only if all non-relation keywords were mapped correctly by the system. Pipeline's performance improved with \system\ most notably for KW, with a 96\%, 50\%, and 110\% increase for MAS, Yelp, and IMDB respectively. The improvement on NaLIR was 5\% for MAS, 13\% for Yelp, and 42\% for IMDB.

\subsubsection{Join Path Inference}

In Table~\ref{tab:logjoin}, we investigate the effect of the Join Path Generator. We focus on Pipeline+ for space reasons, and because improvement was not as drastically evident in NaLIR for reasons described in Section~\ref{sec:evaldisc}.

Activating the Join Path Generator (LogJoin ``Y'') increased accuracy by 11\% for MAS, 24\% for Yelp, and 6\% for IMDB. The combined effect of this with the Keyword Mapper enabled the overall improvement through \system.

\subsection{Error Analysis}
\label{sec:evaldisc}

Augmenting Pipeline with \system\ had a more dramatic effect than with NaLIR because it was given perfectly parsed keywords and metadata as input. Pipeline consequently had a much higher ceiling for improvement compared to NaLIR. While NaLIR is designed to be able to return the relevant metadata, in practice, the system's parser had trouble digesting the correct metadata from NLQs with explicit relation references, such as the token {\em papers} in {\em Return the authors who have papers in Conference X} for MAS, or other NLQs which resulted in nested subqueries. Our takeaway from this is that NLIDBs with better parsers will reap greater benefits from \system, and are hopeful as off-the-shelf parsers have drastically improved since NaLIR's original release.

\subsection{Impact of Parameters}

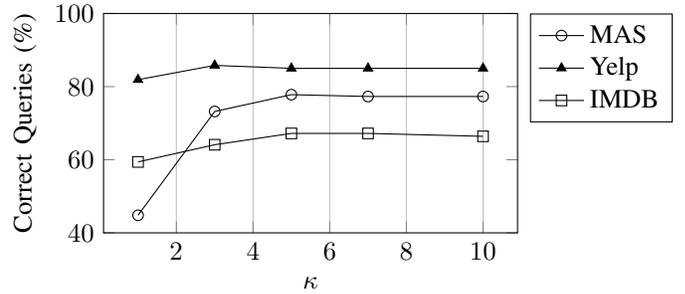
\begin{figure}[t]
\begin{tikzpicture}
\begin{axis}[
    width=0.8\columnwidth,
    height=4.5cm,
    xlabel={$\kappa$},
    ylabel={Correct Queries (\%)},
    xmajorgrids=true,
    ymin=40, ymax=100,
    legend cell align=left,
    legend pos=outer north east
    ]

	\addplot[color=black, mark=o]
coordinates {(1,44.8)(3,73.2)(5,77.8)(7,77.3)(10,77.3)};

	\addplot[color=black, mark=triangle*]
coordinates {(1,81.9)(3,85.8)(5,85.0)(7,85.0)(10,85.0)};

	\addplot[color=black, mark=square]
coordinates {(1,59.4)(3,64.1)(5,67.2)(7,67.2)(10,66.4)};

\legend{MAS, Yelp, IMDB}
\end{axis}
\end{tikzpicture}

\vspace{-0.3cm}

\caption{
  Accuracy of Pipeline+ on each benchmark given a value of $\kappa$, with $\lambda$ fixed at 0.8.
}

\label{fig:kappaacc}
\end{figure}

\begin{figure}[t]
\begin{tikzpicture}
\begin{axis}[
    width=0.8\columnwidth,
    height=4.5cm,
    xlabel={$\lambda$},
    ylabel={Correct Queries (\%)},
    xmajorgrids=true,
    ymin=20, ymax=100,
    xmin=0.0, xmax=1.0,
    legend cell align=left,
    legend pos=outer north east
    ]

	\addplot[color=black, mark=o]
coordinates {(0,76.3)(0.1,78.4)(0.25,78.4)(0.4,78.4)(0.5,77.8)(0.625,79.4)(0.75,79.4)(0.8,77.8)(0.875,69.1)(0.9,63.9)(0.95,51.5)(0.99,45.4)(1,39.7)};

	\addplot[color=black, mark=triangle*]
coordinates {(0,66.9)(0.1,82.7)(0.25,84.3)(0.4,83.5)(0.5,85.0)(0.625,84.3)(0.75,85.0)(0.8,85.0)(0.875,85.8)(0.9,85.8)(0.95,85.0)(0.99,85.0)(1,56.7)};

	\addplot[color=black, mark=square]
coordinates {(0,57.8)(0.1,60.2)(0.25,61.7)(0.4,63.3)(0.5,63.3)(0.625,64.8)(0.75,67.2)(0.8,67.2)(0.875,67.2)(0.9,67.2)(0.95,68.8)(0.99,61.7)(1,32.0)};

\legend{MAS, Yelp, IMDB}
\end{axis}
\end{tikzpicture}

\vspace{-0.3cm}

\caption{
  Accuracy of Pipeline+ on each benchmark given a value of $\lambda$, with $\kappa$ fixed at 5.
}

\label{fig:lambda}
\vspace{-0.5cm}
\end{figure}
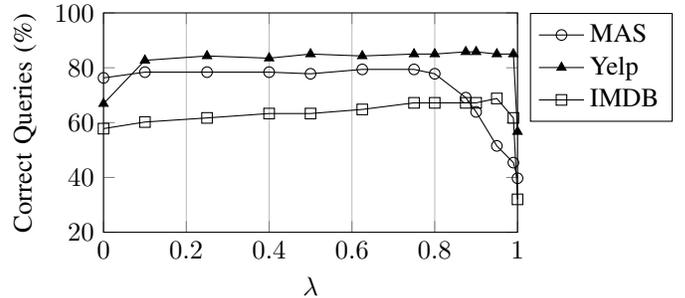

In addition to the system options, there are two parameters that are required to be set in \system: $\kappa$ and $\lambda$. $\kappa$ is the number of top candidate keyword mappings to return before generating configurations, and $\lambda$ is the weight given to the word similarity score as opposed to the log-driven score. We observed the effects of these parameters on Pipeline+.

Figure~\ref{fig:kappaacc} shows that any $\kappa \geq 5$ yields more or less consistent performance. Consequently, we chose $\kappa = 5$ as a cutoff for all our benchmarks because it reflected optimal performance and queries were also evaluated in a timely manner.

In addition, we evaluated the end-to-end performance of Pipeline+ with varying values of $\lambda$ and find similar performance across all benchmarks for $0.1 \leq \lambda \leq 0.8$. For the Yelp benchmark, accuracy falls when $\lambda$ is 0 because the word similarity scores are necessary when ranking configurations, while for the other benchmarks, the pruning procedure for candidate mappings is sufficient to retain and distinguish the correct mappings. Accuracy gradually drops on the MAS and IMDB benchmarks for $\lambda > 0.8$, and sharply on all benchmarks as $\lambda$ approaches 1, suggesting that the log information is crucial for most queries.

\section{Conclusion}
\label{sec:concl}

In this paper, we have described \system, a system that enhances the performance of existing NLIDBs using SQL query logs. We model the information in the SQL query log in a data structure called the Query Fragment Graph, and use this information to improve the ability of existing NLIDBs to perform keyword mapping and join path inference. 
We demonstrated a significant improvement in accuracy when augmenting existing pipeline NLIDBs using log information with \system.  Possible future work includes exploring the influence of user sessions in the SQL query log, as well as finding ways to improve existing deep learning-based end-to-end NLIDB systems with information from the SQL log.

\section{Acknowledgments}

This work was supported in part by IBM under contract 4915012629 and by NSF grant IIS-1250880.

\bibliographystyle{abbrv}
\bibliography{paper}

\end{document}